\documentclass[preprint,12pt]{article}
 
\usepackage{amssymb}
\usepackage{caption}
\usepackage{breqn}

\begin{document}

\textbf{BFKL Eigenvalue  and Maximal Alternation of Harmonic Sums}
\begin{center}

\small
Alex Prygarin
\\
Department of Physics, Ariel University, Ariel 40700, Israel 
\end{center}
\normalsize

\normalsize

\begin{abstract}
We analyze the known results for the eigenvalue of the Balitsky-Fadin-Kuraev-Lipatov~(BFKL) equation in the perturbative regime using the analytic continuation of   harmonic sums from even positive  arguments to the complex plane. The resulting meromorphic functions have poles at negative integer values of the argument. 
The typical classification of harmonic sums is determined by two major parameters: $a)$ the \textit{weight} - a sum of inverse powers of the summation indices; 
$b)$ the  \textit{depth} - a number of nested summations. We introduce the  third parameter: the \textit{alternation} - a number of nested sign-alternating summations in a given harmonic sum.  We claim that the maximal alternation of the nested summation in the functions building the BFKL eigenvalue is preserved from loop to loop in the perturbative expansion. 
The BFKL equation is formulated for arbitrary color configuration of the propagating states in the $t$-channel. Based on known results one can state that color  adjoint BFKL eigenvalue be can written using only harmonic sums with positive indices, maximal alternation zero,  and at most depth one, whereas the singlet BFKL eigenvalue is constructed of harmonic sums with maximal sign alternation being equal one. We also note that for  maximal alternation being equal  unity the harmonic sums can be expressed through alternation zero harmonic sums with half-shifted arguments.   

\end{abstract}

\section{Introduction}\label{}

The Balitsky-Fadin-Kuraev-Lipatov~(BFKL)~\cite{BFKL} approach is used to  resum leading logarithms of energy  in the framework of the perturbative approach to the gauge field theories. Two most useful and famous examples of the BFKL approach are the BFKL equation in QCD and  maximally supersymmetric Yang-Mills Theory~($N=4$ SUSY) that describe the propagation of two reggeized gluons in the limit of transferred momenta being much smaller the center-of-mass energy. The QCD version of the BFKL equation was shown to comply with experimental data, while its supersymmetric extension was found to be very useful in predicting the high energy behavior of helicity amplitudes. The BFKL equation was originally formulated for an arbitrary color state propagating in the $t$-channel having a universal structure for all scattering particles due to Regge factorization. This fact allowed to isolate the high energy evolution from the impact factors and consider it in separate. Another  fascinating property of the BFKL equation is the  bootstrap, which was used to check the self-consistency of the approach~\footnote{For more details on the use of the bootstrap, Regge factorization and the proof  of the gluon  reggeization  in the framework of the perturbation theory the reader is referred to a profound  review book by Ioffe, Fadin and Lipatov~\cite{Ioffe:2010zz} and references wherein. }.

The original BFKL equation was derived four decades ago in the leading order and it took another decade to find its exact solution in the color singlet state exploiting the conformal symmetry of the resulting Hamiltonian~\cite{int}. The next-to-leading logarithmic correction to the BFKL equation calculated shortly after that also was shown to enjoy the same conformal symmetry  in the color singlet state and was eventually solved using the LO eigenfunctions. 
It is worth mentioning that the LO BFKL in QCD and $N=4$ SUSY coincide because in the LO only states propagating in the $t$-channel  with the highest spin~(gluons) contribute and those are identical in QCD and $N=4$ SUSY. The differences between the two starts to show up already at the next-to-leading order, where quarks and gluino both having spin-$\frac{1}{2}$ must be included. 

The main motivation for the BFKL equation was to derive a perturbative representation of the leading Regge trajectory~(Pomeron) in the framework of the  gauge theory, mainly in QCD. As such most of attention was attracted to the  color singlet representation, while the color adjoint state was used only to show the self consistency of the approach and demonstrate the gluon reggeization. The color adjoint BFKL state collapses to the propagation of the $t$-channel reggeized gluon if two $t$-channel reggeized gluons have the same  particle-reggeon-particle transitions vertices. 
 Bartles, Lipatov and Sabio Vera~\cite{BLS1, BLS2}  considered other, reggeon-particle-reggeon transitions vertices  in attempt to analyze the analytical structure of the planar scattering amplitudes with definite helicity configurations of the external particles in $N=4$ SUSY.
In this case the  adjoint BFKL state does not reduce to the reggeized gluon and led to introduction of the BFKL equation in the color adjoint state.     In contrast to the infrared safe color singlet BFKL, in the color adjoint case the infrared singularities does not cancel and typically removed by \textit{ad hoc}   procedure where they are absorbed into the Bern-Dixon-Smirnov~(BDS)~\cite{BDS} amplitude. This uncertainty in removing the infrared divergent part makes the notation of color adjoint BFKL equation rather vague allowing to numerous  prescriptions for removing infrared part. Apart from the commonly used the BDS-like "regularization" of the color adjoint BFKL equation, where one removes twice the infrared piece of  one of the  reggeized gluons, there exists a more symmetric recipe in which one can remove half of each of two gluons on equal basis. To the best of our knowledge nobody considered this possibility despite the fact that it is a more natural way to remove the infrared divergent pieces. In the present discussion we are focused only on the commonly used BDS-like prescription for removing infrared divergent parts of the BFKL equation projected on the color adjoint state of the reggeized gluons in the $t$-channel. In this regularization prescription the resulting final part builds Hamiltonian  that has  conformal symmetry in the dual momentum space. This is in contrast to the  color singlet BFKL that enjoys the conformal symmetry in the coordinate space.

 The conformal symmetry was used to find the exact solution of the BFKL equation in the color adjoint representation to any order of the perturbative expansion and also extract the leading strong coupling behavior. In contrast to that, the situation with the color singlet BFKL is more complicated and at the moment we know only the leading and the next-to-leading order solution to the BFKL in both QCD and $N=4$ SUSY. There are also recent attempts to find the  BFKL eigenvalue in the color singlet representation in $N=4$ SUSY at the higher orders using integrability techniques by Gromov, Levkovich-Maslyuk and Sizov~\cite{Gromov:2015vua}, Caron Huot and Herraren~\cite{Caron-Huot:2016tzz} and  Alfimov, Gromov and Sizov~\cite{Alfimov:2018cms}. The main outcome of   this analysis is a closed system of equation that allows a rather precise numerical solution, but still has to be solved in terms of known functions for arbitrary values of anomalous dimension and conformal spin. In this paper we consider the   main results of the integrability approaches to the color singlet BFKL with the focus of use of the harmonic sums of one variable as a potential candidate for a space of functions building the BFKL eigenvalue at any order of the perturbation theory.

 \section{Harmonic Sums and BFKL Eigenvalue.}

The BFKL equation can be schematically written as the stationary  Schroedinger equation 
\begin{eqnarray}
H \Psi_{\nu, n} = E_{\nu, n} \Psi_{\nu, n},
\end{eqnarray}  
where $\Psi_{\nu, n} $ are the eigenfunctions corresponding to the eigenvalues $E_{\nu, n}$ that we call the BFKL energy. 
The Hamiltonian $H$ is different for different color representations and gauge theory under discussion, but it shares the common feature of integrability~\cite{int}. 
The integrability appears mostly due to the fact that Hamiltonian is formulated in two dimensional space of  the transverse transferred momentum, where it enjoys the conformal symmetry in either coordinate or momentum space for color singlet to color adjoint representations respectively. 
The  BFKL eigenfunctions are built of conformal cross ratios in coordinate space for singlet and in the dual transverse momentum space for adjoint color configuration. In order to find the eigenvalue it is enough to consider the eigenfunctions for zero transferred momentum 
 $\Psi_{\nu, n} \simeq x^{\frac{1}{2}+ i \nu +\frac{n}{2}}\bar{x}^{\frac{1}{2}+ i \nu -\frac{n}{2}}$ for color singlet BFKL and 
  $\Psi_{\nu, n} \simeq k^{ i \nu +\frac{n}{2}}\bar{k}^{i \nu -\frac{n}{2}}$  for color adjoint BFKL equation, where $\nu$ is the  anomalous dimension and $n$ is the  conformal spin that takes integer values $n=0, \pm 1, \pm 2 ,...$.

The BFKL eigenvalue is given by the perturbative expansion in the coupling constant $a= \frac{g_{YM}^2 N_c}{ 8 \pi^2}$, where $N_c$ is the number of colors and reads
\begin{eqnarray}
E_{\nu,n}= a \left( E^{LO}_{\nu,n}+ a E^{NLO}_{\nu,n}+a^2 E^{NNLO}_{\nu,n}+...\right)
\end{eqnarray}

The leading order~(LO) eigenvalue $E^{LO}_{\nu,n}$ for both singlet and adjoint cases can be written using only digamma function 
$\psi(z)= \frac{d \ln \Gamma(z)}{d z}$. 
For the singlet case $E^{LO,S}_{\nu,n}$ is given by 
\begin{eqnarray}\label{LOsinglet}
E^{LO,S}_{\nu,n} &=&-\frac{1}{2}  \psi\left(-\frac{1}{2}+i \nu +\frac{n}{2}\right)- \frac{1}{2} \psi\left(-\frac{1}{2}-i \nu +\frac{n}{2}\right)
\nonumber
\\
&&-\frac{1}{2} \psi\left(-\frac{1}{2}+i \nu -\frac{n}{2}\right)-\frac{1}{2} \psi\left(-\frac{1}{2}-i \nu -\frac{n}{2}\right)+ 2\psi\left(1\right) 
\end{eqnarray}
while for the color adjoint BFKL it reads
\begin{dmath}\label{LOadjoint}
E^{LO,A}_{\nu,n} = \frac{1}{2} \psi\left(i \nu +\frac{n}{2}\right)+\frac{1}{2}\psi\left(-i \nu +\frac{n}{2}\right)+\frac{1}{2}\psi\left(+i \nu -\frac{n}{2}\right)+\frac{1}{2}\psi\left(-i \nu -\frac{n}{2}\right)-2 \psi\left(1\right)  
\end{dmath}
The two expressions look very similar however they are quite different in the following aspects. Firstly, the meaning of the anomalous dimension $\nu$ and the conformal spin $n$ is different because they are defined for different conformal symmetries, the conformal symmetry in the coordinates space for $E^{LO,S}_{\nu,n}$ and the conformal symmetry in the dual momentum space for $E^{LO,A}_{\nu,n}$. Secondly, $E^{LO,A}_{\nu,n}$ goes to zero as $\nu \to 0$ for $n=0$, while $E^{LO,S}_{\nu,n}$ takes a finite value in this limit that corresponds to BFKL Pomeron intercept and defines the leading energy behavior. Not less important limit  is $\nu \to 0$ for $|n|=1$, which corresponds to overlap of the Regge kinematics with the collinear kinematics. In this kinematics BFKL and DGLAP equations give the same result. One can see that for  $\nu \to 0$  and $|n|=1$ the singlet eigenvalue $E^{LO,S}_{\nu,n}$ vanishes, while $E^{LO,A}_{\nu,n}$ is divergent in this limit. 

Despite the major differences in the analytic behavior and the physical meaning $E^{LO,S}_{\nu,n}$ and $E^{LO,A}_{\nu,n}$ are still very similar and one can speculate about direct transition between the two expressions skipping going back to the full BFKL equation for arbitrary number of colors that must be projected on the two color states in order to eventually have  $E^{LO,S}_{\nu,n}$ and $E^{LO,A}_{\nu,n}$~\cite{Bondarenko:2015tba,Bondarenko:2016tws}. Naively this transition at the LO level can be done by a simple shift of 
the argument by $\frac{1}{2}$. Technically this difference in $\frac{1}{2}$ of the argument originates from the different normalization conditions of the wave functions in the coordinate and the momentum spaces. A more profound explanation would be a different representation in terms of the Heisenberg spin chain, the color singlet BFKL is associated with closed spin chain, while the BFKL in the adjoint representation corresponds to the open spin chain that implies the different boundary condition for the two cases. 

  The expressions $E^{LO,S}_{\nu,n}$ and  $E^{LO,A}_{\nu,n}$  in   QCD and $N=4$ SUSY coincide  because only particles with the highest spin, i.e. gluons, contribute to the leading logarithm term. Already at the NLO level there is a difference between QCD and $N=4$ SUSY due to contributions from quarks and gluinos. In this paper we limit our discussion to $N=4$ SUSY mainly due the fact that the BFKL eigenvalue is currently known only to the next-to-leading order in QCD for both color singlet and adjoint cases, while in  $N=4$ SUSY the adjoint BFKL value is known to any order and we have a lot of information about the singlet BFKL eigenvalue at higher orders in some particular limits. 

We have mentioned that one can naively formulate the transition between the singlet and the adjoint BFKL eigenvalues as a simple "shift of the argument" by one half. The relation between singlet and adjoint cases is even less obvious already at the next-to-leading order where $E^{NLO,A}_{\nu,n}$  in $\mathcal{N}=4$ SYM can be expressed merely in 
terms of the polygamma functions~( derivatives of digamma function) and ratio functions~\cite{Fadin:2011we}
\begin{eqnarray}
 E^{NLO,A}_{\nu,n}&=&-\frac{1}{4}\left(\psi ^{\prime \prime}(1+i\nu +\frac{|n|}{2})+
\psi ^{\prime \prime}(1-i\nu +\frac{|n|}{2})
\right. 
\nonumber
\\
&&
\left. +\frac{2i\nu \left(\psi '(1-i\nu +\frac{|n|}{2})-\psi'(1+i\nu
+\frac{|n|}{2})\right)}{\nu ^2+\frac{n^2}{4}}
\right)
 \nonumber
\\
&&
-\zeta (2)\, E^{LO,A}_{\nu,n}-3\zeta (3)-\frac{1}{4}\,\frac{|n|\,\left(\nu
^2-\frac{n^2}{4}\right)}{\left(\nu
^2+\frac{n^2}{4}\right)^3}\,,
\label{NLOadjoint}
\end{eqnarray}
whereas $E^{NLO,S}_{\nu,n}$ is more complicated and  includes new kind of functions, which are generalization of  Lerch zeta function 
\begin{eqnarray}
L(\lambda, \alpha,s )=\sum_{k=0}^\infty \frac{e^{i 2 \pi\lambda k }}{(k+\alpha)^s}.
\end{eqnarray} 

The NLO singlet eigenvalue $E^{NLO,S}_{\nu,n}$  in $\mathcal{N}=4$ SYM reads~\cite{Fadin:1998py,Kotikov:2001sc,Kotikov:2002ab}
\begin{eqnarray}\label{NLOsinglet}
E^{NLO,S}_{\nu,n}&=& \Phi\left(\frac{1}{2}+i \nu +\frac{|n|}{2}\right)+\Phi\left(\frac{1}{2}-i \nu +\frac{|n|}{2}\right) \nonumber
\\
&&
-\frac{1}{2}E^{LO,S}_{\nu,n} \left( \beta' \left(\frac{1}{2}+i \nu +\frac{|n|}{2}\right)+\beta' \left(\frac{1}{2}-i \nu +\frac{|n|}{2}\right) +\zeta(2)\right),
\end{eqnarray}
where~\cite{handbuch}
\begin{eqnarray}
\beta'(z)=\sum_{r=0}^{\infty} \frac{(-1)^{r+1}}{(z+r)^2}
\end{eqnarray}
and 
\begin{eqnarray}\label{phi}
\Phi(z)= 3 \zeta(3) +\psi^{''} (z) +2 \Phi_2(z)+2 \beta'(z) \left(\psi(1)-\psi(z)\right). 
\end{eqnarray}
The function $\Phi_2(z)$ is the most complicated function announced above, it is defined by 
\begin{eqnarray}
\Phi_2(z)= \sum_{k=0}^{\infty} \frac{\beta'(k+1)+ (-1)^k \psi'(k+1)}{k+z}-\sum_{k=0}^{\infty}\frac{(-1)^k (\psi(k+1)-\psi(1))}{(k+z)^2}
\end{eqnarray}
and related to the generalized harmonic sums.
The  complexity of the function $\Phi_2(z)$ is determined by two things, by its transcendentality~(equal to three)~\footnote{The \textit{ad hoc}  definition of the transcendentality of $\Phi_2(z)$ is based on the fact that it gives a linear combination of  constants of uniform transcendentality as $z\to \infty$.  }
 and the number of the nested summations~(equal to one).
The main question which is still to be answered what are the functions that build higher order corrections to the singlet BFKL eigenvalue. Its is quite clear that 
the functions entering the higher order singlet BFKL eigenvalue must have higher transcendentality and more nested summations.  There is another complexity parameter that was overlooked in the literature for some reason- the maximal sign alternation.   One can see that the maximal sign alternation of each term of the  function $\Phi_2(z)$ equals one, which means that only one of two nested summations is the sign alternating summation.  Most probably, the maximal sign alternation being equal one is the artifact of the integrability of the closed spin chain.

\subsection{Harmonic Sums}
The harmonic sums are defined through a nested summation with their argument being the upper limit in the outermost sum~\cite{HS1,Vermaseren:1998uu}
\begin{eqnarray}\label{defS}
S_{a_1,a_2,...,a_k}(n)=  \sum_{n \geq i_1 \geq i_2 \geq ... \geq i_k \geq 1 }   \frac{\mathtt{sign}(a_1)^{i_1}}{i_1^{|a_1|}}... \frac{\mathtt{sign}(a_k)^{i_k}}{i_k^{|a_k|}}, \;\; n \in \mathbb{N}^*
\end{eqnarray}
In this paper we consider the harmonic sums with only real integer values of $a_i$, which build the alphabet of the possible negative and positive indices.  
In Eq.~(\ref{defS}) $k$ is  the depth and $w=\sum_{i=1}^{k}|a_i|$ is the weight of the harmonic sum $S_{a_1,a_2,...,a_k}(n)$. 

The indices of harmonic sums $a_1,a_2,...,a_k$ can be either positive or negative integers and label uniquely $S_{a_1,a_2,...,a_k}(n)$ for any given 
weight. However there is no unique way of building the functional basis for a given weight because the harmonic sums are subject to so called quasi-shuffle relations, where a linear combination of $S_{a_1,a_2,...,a_k}(n)$ with the same argument but all possible permutations of indices can be expressed through 
a non-linear combinations of harmonic sums at lower weight.
 There is also some freedom in choosing the irreducible minimal set of $S_{a_1,a_2,...,a_k}(n)$ that builds those non-linear combinations.  
The quasi-shuffle relations make a connection between the linear and non-linear combinations of the harmonic sums of the same argument. For example, 
 at depth two the quasi-shuffle relation reads
\begin{eqnarray}\label{shuffle}
S_{a,b}(z)+S_{b,a}(z)= S_{a}(z) S_{b}(z)+S_{\textrm{ sign}(a) \textrm{ sign}(b)(|a|+|b|)}(z)
\end{eqnarray}
The quasi-shuffle relations of the harmonic sums are closely connected to the algebra of the harmonic polylogarithms~\cite{Vermaseren:1998uu}.

There is another type of identity called the duplication identities where a combination of harmonic sums of argument $n$ can be expressed through 
a harmonic sum of the argument $2 n$. The duplication identities introduce another freedom in choosing the functional basis. 

  In this paper we consider the analytic continuation of the harmonic sums from positive integer values of the argument to the complex plane denoted by $\bar{S}_{a_1,a_2,...}^{+}(z)$~(this notation was introduced by Kotikov and Velizhanin~\cite{Kotikov:2005gr}). The analytic continuation is done in terms of the Mellin transform of corresponding Harmonic Polylogarithms and was recently used by Gromov, Levkovich-Maslyuk and Sizov~\cite{Gromov:2015vua,Alfimov:2018cms}~\footnote{See also work of  Velizhanin~\cite{Velizhanin:2015xsa}} and Caron Huot and Herraren~\cite{Caron-Huot:2016tzz} for expressing the  eigenvalue of the Balitsky-Fadin-Kuraev-Lipatov~(BFKL)  equation using the principle of Maximal Transcedentality~\cite{Kotikov:2006ts} in super Yang-Mills $\mathcal{N}=4$ field theory. We plan to use their results together with analysis done by one of the authors and collaborators~\cite{Bondarenko:2015tba,Bondarenko:2016tws,Prygarin:2018tng,Prygarin:2018cog} to understand the general structure of the   BFKL equation in QCD and beyond.

 The Mellin transform allows to make the analytic continuation to the complex plane. For example, consider the harmonic sum
 that is typically used in the literature to demonstrate the procedure of the analytic continuation 
 \begin{eqnarray}\label{s1}
 S_{-1}(z)=\sum_{k=1}^{z} \frac{(-1)^k}{k}, \;\; z \in \mathbb{N}^*
\end{eqnarray}  
The corresponding Mellin transform reads
\begin{eqnarray}
\int_0^1 \frac{1}{1+x} x^z= (-1)^z \left(S_{-1}(z)+\ln 2\right)
\end{eqnarray}
One can see that $S_{-1}(z)$ on its own is not an analytic function because of the term $(-1)^z $ and we impose that we start from even integer values of the argument $z$. In this case we define its analytic continuation from even positive integers to all positive integers through 
\begin{eqnarray}
\bar{S}_{-1}^{+}(z)=(-1)^z S_{-1}(z)+((-1)^z-1) \ln 2
\end{eqnarray} 
 and thus we can write 
 \begin{eqnarray}
 \bar{S}_{-1}^{+}(z)=\int_0^1 \frac{1}{1+x} x^z-\ln 2
\end{eqnarray}
This way we defined  $\bar{S}_{-1}^{+}(z)$ using the Mellin transform of ratio function $ \frac{1}{1+x}$. In more complicated cases of other harmonic sums one includes also Harmonic Polylogarithms on top of the ratio functions, but the general procedure is very similar and largely covered in  a number of publications~(See for example the paper of Kotikov and Velizhanin~\cite{Kotikov:2005gr} or very detailed Thesis of Alblinger~\cite{AblingerThesis,Ablinger:2011te}).

 It is worth mentioning that there is another analytic continuation for the harmonic sum, from odd positive integer values of the argument, which is different for harmonic sums with at least one negative index and  denoted by  $\bar{S}_{a_1,a_2,...}^{-}(z)$. 
Both analytic continuations are equally valid. 

In this paper we use only analytic continuation of the harmonic sums from even positive integers to the complex plane, namely,  $\bar{S}_{a_1,a_2,...}^{+}(z)$ and for the clarity of presentation  we denote $\bar{S}_{a_1,a_2,...}^{+}(z)$ by simply $S_{a_1,a_2,...}(z)$ throughout the text below.

\subsection{LO and NLO BFKL Eigenvalue through Harmonic Sums}
In this section we write the leading and next-to-leading BFKL eigenvalues in the color singlet and color adjoint representations in terms of the harmonic sums analytically continued from the positive even integer argument to the complex plane. For brevity we omit the \emph{bar-plus} notation of $\bar{S}_{a_1,a_2,...}^{+}(z)$ and use $S_{a_1,a_2,...}(z)$ instead. 

The leading order BFKL in eq.~(\ref{LOsinglet}) and eq.~(\ref{LOadjoint}) read

\begin{eqnarray}\label{LOsingletS}
E^{LO,S}_{\nu,n} &=&-\frac{1}{2} S_1\left(\frac{1}{2}+i \nu +\frac{n}{2}\right)- \frac{1}{2} S_1\left(\frac{1}{2}-i \nu +\frac{n}{2}\right)
\nonumber
\\
&& -\frac{1}{2} S_1\left(\frac{1}{2}+i \nu -\frac{n}{2}\right)-\frac{1}{2} S_1\left(\frac{1}{2}-i \nu -\frac{n}{2}\right)
\end{eqnarray}
while for the color adjoint BFKL it reads
\begin{eqnarray}\label{LOadjointS}
E^{LO,A}_{\nu,n} &=& \frac{1}{2} S_1\left(-1+i \nu +\frac{n}{2}\right)+\frac{1}{2}S_1\left(-1-i \nu +\frac{n}{2}\right)
\nonumber
\\
&&
+\frac{1}{2}S_1\left(-1+i \nu -\frac{n}{2}\right)+\frac{1}{2}S_1\left(-1-i \nu -\frac{n}{2}\right)  
\end{eqnarray}
where we used $\psi(1+z)-\psi(1)=S_1(z)$.

The next-to-leading color singlet eigenvalue $E^{NLO,S}_{\nu,n}$ ca be written as 
\begin{eqnarray}\label{NLOsingletS}
E^{NLO,S}_{\nu,n}&=& \Phi\left(\frac{1}{2}+i \nu +\frac{|n|}{2}\right)+\Phi\left(\frac{1}{2}-i \nu +\frac{|n|}{2}\right) \nonumber
\\
&&
-\frac{1}{2}E^{LO,S}_{\nu,n} \left( S_{-2} \left(-\frac{1}{2}+i \nu +\frac{|n|}{2}\right)+S_{-2} \left(-\frac{1}{2}-i \nu +\frac{|n|}{2}\right) +2\zeta(2)\right),\nonumber 
\end{eqnarray}
where 
\begin{eqnarray}\label{PhitoS}
\Phi(z)=4 S_{1,-2}(z-1)-2 S_{-3}(z-1)  +2 S_3(z-1)+\frac{\pi^2}{3} S_1 (z-1) 
\end{eqnarray}
Here we used 
\begin{eqnarray}
\beta'(z)= S_{-2}(z-1) +\frac{\zeta(2)}{2}
\end{eqnarray}
and some lengthy calculation   relating  the functions appearing in eq.~(\ref{phi}) to $S_{a_1, a_2}(z-1)$

In a similar way we can write the color adjoint BFKL eigenvalue $ E^{NLO,A}_{\nu,n}$ in eq.~(\ref{NLOadjoint}) in terms of the harmonic sums. Firstly we note that the ratio function can be written as a linear combination of the polygamma functions with shifted argument as follows
\begin{eqnarray}
&&\frac{2 i \nu }{\nu^2+\frac{n^2}{4}}= \left\{-\psi\left(1+ i \nu +\frac{|n|}{2}\right)+\psi\left( i \nu +\frac{|n|}{2}\right)
\right.
\nonumber 
\\
&&
\left.
+\psi\left(1- i \nu +\frac{|n|}{2}\right)-\psi\left(-i \nu +\frac{|n|}{2}\right)\right\}
\end{eqnarray}
and   
\begin{small}
\begin{eqnarray}
-\frac{1}{4}\frac{|n|\left(\nu^2-\frac{n^2}{4}\right)}{\left(\nu^2+\frac{n^2}{4}\right)^3}&=&
 -\frac{1}{4} \left\{\psi'\left(1+ i \nu +\frac{|n|}{2}\right)-\psi'\left( i \nu +\frac{|n|}{2}\right)
 \right.
\nonumber 
\\
&&
\left.
+\psi'\left(1- i \nu +\frac{|n|}{2}\right)-\psi'\left(-i \nu +\frac{|n|}{2}\right)\right\} \nonumber
 \\
 &&\times 
 \left\{\psi\left(1+ i \nu +\frac{|n|}{2}\right)-\psi\left( i \nu +\frac{|n|}{2}\right)
 \right.
\nonumber 
\\
&&
\left.
+\psi\left(1- i \nu +\frac{|n|}{2}\right)-\psi\left(-i \nu +\frac{|n|}{2}\right)\right\}
\end{eqnarray}
\end{small}
Then we expressed each of the polygamma function in terms of harmonic sums with the shifted argument
\begin{eqnarray}\label{rational1}
\frac{2 i \nu }{\nu^2+\frac{n^2}{4}}&=& \left\{-S_1\left( i \nu +\frac{|n|}{2}\right)+S_1\left(-1+ i \nu +\frac{|n|}{2}\right)
\right.
\nonumber 
\\
&&
\left.
+S_1\left(- i \nu +\frac{|n|}{2}\right)-S_1\left(-1-i \nu +\frac{|n|}{2}\right)\right\}
\end{eqnarray}
and 
\begin{small}
\begin{eqnarray}\label{rational2}
-\frac{1}{4}\frac{|n|\left(\nu^2-\frac{n^2}{4}\right)}{\left(\nu^2+\frac{n^2}{4}\right)^3}&=&
 -\frac{1}{4} \left\{S_2\left( i \nu +\frac{|n|}{2}\right)-S_2\left(-1+ i \nu +\frac{|n|}{2}\right)
 \right.
\nonumber 
\\
&&
\left.
+S_2\left(- i \nu +\frac{|n|}{2}\right)-S_2\left(-1-i \nu +\frac{|n|}{2}\right)\right\} \nonumber
 \\
 &&\times 
 \left\{S_1\left(i \nu +\frac{|n|}{2}\right)-S_1\left(-1+ i \nu +\frac{|n|}{2}\right)
 \right.
\nonumber 
\\
&&
\left.
+S_1\left(- i \nu +\frac{|n|}{2}\right)-S_1\left(-1-i \nu +\frac{|n|}{2}\right)\right\}, 
\end{eqnarray}
\end{small}
where we used  $-\psi'(z+1)+\psi'(1)=S_{2}(z)$.

Finally, we write
\begin{eqnarray}
 && E^{NLO,A}_{\nu,n}= -\frac{1}{2} S_3(z)-\frac{1}{2} S_3(\bar{z}) -2 \zeta(3) -\zeta(2)  E^{NLO,A}_{\nu,n} \nonumber
 \\
 &&
+ \left\{ 
 -S_1(z) +S_1(z-1)+S_1(\bar{z})-S_1(\bar{z}-1)  
 \right\}
 \times 
 \left\{
 S_2(z)-S_2(\bar{z})
 \right\},
\label{NLOadjointS}
\end{eqnarray}
using $\psi''(z+1)= 2 S_3 (z) -2 \zeta(3)$.

The obvious observation is that the BFKL eigenvalue in $N=4$ SUSY can be written only in terms of the harmonic sums analytically continued to the complex plane~\footnote{Strictly speaking, by analytically continued harmonic sum we mean a corresponding meromorphic function defined everywhere on the complex plane except for isolated poles at negative integers}. Due to the principle of maximal transcendentality formulated by Kotikov and Lipatov~\cite{Kotikov:2006ts} the harmonic sums entering each expression are of the same weight, weight $w=1$ for LO and $w=3$ for NLO.
The weight of harmonic sums defines their transcendentality through the transcendentality of the constants obtained from the value of the harmonic sums at infinity. 
 Each further order of the perturbation theory will increase the weight by two, for next-to-next-to-leading order the weight is $w=5$.  In contrast to other function appearing in  eq.~(\ref{phi}) the harmonic sums are easily generalized to higher weight. The major technicality of the harmonic sums is the analytic continuation to the complex plane, which can be done in a straightforward way by introducing Mellin transform of harmonic polylogarithms~(HPL). Another troublesome aspect is that harmonic sums at any weight form overcomplete basis due to quasi-shuffle identities relating a linear combination of harmonic sums to a product of two harmonic sums at lower weight. 
One way of circumfering this issue is to choose the linear basis, for example at weight $w=4$ the linear basis would include only the following pure sums
\begin{eqnarray}
&&\left\{S_{-3},S_3,S_{-2,-1},S_{-2,1},S_{2,-1},S_{2,1},S_{-1,1,-1}
   ,S_{-1,1,1},S_{1,-2},S_{1,2},S_{1,-1,-1}, \nonumber
   \right.
   \\
    && \left.
    S_{1,-1,1},S_{1,1,-1}
   ,S_{1,1,1},  S_{-1,-2},S_{-1,2},S_{-1,-1,-1},S_{-1,-1,1}\right\}.
\end{eqnarray} 

The use of harmonic sums is very useful in building ansatz of the final answer and then implementing integrability techniques to fix the free coefficients.   
In particular, we note that the adjoint BFKL eigenvalue is built of harmonic sums with only positive indices~(maximal alternation zero), whereas the singlet BFKL eigenvalue is built of harmonic sums with at most one negative index~(max alternation one).  This observation is supported by recent NNLO calculations for $n=0$  and for any particular $n$ for singlet case. For the color adjoint BFKL eigenvalue this can be easily shown  to hold at any order  
because it is built of derivatives of polygamma functions and the rational functions, both are expressible through harmonic sums of depth one and  with only positive indices. In order to illustrate the importance of this observation we consider  the total number of harmonic sums at given weight- the second column, the number of harmonic sums with maximal alternation one- the third column, and finally in the last column the number of harmonic sums with only positive indices~(Max alternation zero). 
\newline

\begin{tabular}{ | c |  c | c | c | }
\hline
Weight & Any alternation & Max alternation one & Max alternation zero \\ \hline  
  1 & 2 & 2 & 1 \\ \hline
  2 & 6 & 5  & 2  \\  \hline
  3 & 18 & 12   & 4\\  \hline
  4 & 54 & 28  &  8 \\  \hline
  5 & 162 & 64  &  16 \\  \hline
\end{tabular}
\captionof{table}{The total number of harmonic sums at given weight vs the number of harmonic sums  for maximal alternation one and  zero.}

We see a drastic reduction of the ansatz basis, higher weight more pronounced   becomes this feature.

A further simplification for calculating the singlet BFKL eigenvalue at higher orders may 
the fact 
that the  harmonic sums with at most one negative index can be expressed through harmonic sums with only positive indices and of  half argument shifted by one half. For example consider   two simple harmonic sums, namely two meromrphic functions corresponding to the analytic continuation of $S_2\left(z\right)$ and $S_{-2}\left(z\right)$ from even positive integer arguments   to the complex plane~\footnote{Similar expressions for $\psi'(z)$ and $\beta'(z)$ can be found in the book of Nielsen~\cite{handbuch}.} 
\begin{eqnarray}
S_2\left(z\right)= \frac{1}{4}S_2\left(\frac{z}{2}\right)+ \frac{1}{4}S_2\left(\frac{z}{2}-\frac{1}{2}\right) +\frac{\zeta(2)}{2}
\end{eqnarray}
\begin{eqnarray}
S_{-2}\left(z\right)= \frac{1}{4}S_2\left(\frac{z}{2}\right)- \frac{1}{4}S_2\left(\frac{z}{2}-\frac{1}{2}\right) -\frac{\zeta(2)}{2}
\end{eqnarray}
Here we trade a more complicated basis with one argument for a simpler basis for two possible arguments. Naturally, this procedure does not change the number of free coefficients to be fixed, but brings more insight into the problem under discussion.

\section{Conclusion and Discussions}
We analyze the existing results for the BFKL eigenvalue in $\mathcal{N}=4$ SYM in the adjoint and singlet color representations. The analytic structure in both cases is similar but largely differs by the functions they are built of. This analysis reveals that there is a new conserved quantity that propagates from loop to loop in the perturbative expansion, namely, the sign alternation of nested summation in the functions that build the BFKL eigenvalue. The maximal sign alternation of the functions involved equals one for the color singlet BFKL eigenvalue, while maximal sign alternation equals zero~(sign constant summation) is a multi-loop feature of the  color adjoint BFKL eigenvalue. 

We consider harmonic sums of one argument as possible candidate for a proper space of functions describing the singlet BFKL eigenvalue  at any given order of the perturbation theory.
  It is well known that the harmonic sums of one argument can be continued from positive integer values of the argument to the complex plane using the Mellin transform of the harmonic polylogarithms. The resulting meromorphic functions are well defined at the complex plane except for isolated pole at the 
  negative integers and can used for expressing known results at the leading order~(LO) and the next-to-leading order~(NLO) in $N=4$ SUSY. Same can be done also for QCD, where one includes also harmonic sums at lower weight. The known results for the  BFKL eigenvalue in the color singlet state suggest that it is  built of harmonic sums with at most one negative index~(max alternation one). This fact allows to sort the sums at a given weight by sign alternation removing all sums with alternation bigger than one. This simple procedure drastically reduces the expansion basis and thus allows to fix all free coefficients  with less amount of data. 

This work is dedicated to the memory of Academician  Professor Lev Nikolaevich Lipatov and is based on most exciting and unforgettable discussions with him on the BFKL physics.

\end{document}